\documentclass[letterpaper, 10pt, conference]{ieeeconf}
\IEEEoverridecommandlockouts

\let\labelindent\relax

\usepackage{amsmath,amsfonts}
\usepackage{algorithmic}
\usepackage{algorithm}
\usepackage{array}
\usepackage[caption=false,font=normalsize,labelfont=sf,textfont=sf]{subfig}
\usepackage{textcomp}
\usepackage{stfloats}
\usepackage{url}
\usepackage{verbatim}
\usepackage{graphicx}
\usepackage{cite}
\usepackage{booktabs}
\usepackage{cite}
\usepackage{amsthm}
\usepackage{algorithmic}
\usepackage{graphicx}
\usepackage{textcomp}
\usepackage{xcolor}
\usepackage{graphicx}
\usepackage{indentfirst}
\usepackage{CJK}
\usepackage{graphicx}
\usepackage{epstopdf}
\usepackage{flushend}
\usepackage{balance}
\usepackage{lettrine}
\usepackage{flushend,cuted}
\usepackage{blindtext}
\usepackage[caption=false,font=normalsize,labelfont=sf,textfont=sf]{subfig}
\usepackage{amsmath,amsfonts}
\usepackage{algorithm}
\usepackage{array}
\usepackage{enumerate}
\usepackage{enumitem} 

\usepackage{orcidlink}

\newtheorem{problem}{Problem}
\newtheorem{theorem}{Theorem}
\newtheorem{lemma}{Lemma}
\newtheorem{assumption}{Assumption}

\begin{document}
	
	\title{\LARGE \bf Fixed-Relative-Switched Threshold Strategies for Consensus Tracking Control of Nonlinear Multiagent Systems}


	\author{Ziming Wang$^{1}$\orcidlink{0000-0001-7000-9578}, Yun Gao$^{1}$\orcidlink{0000-0002-3061-8595}, Apostolos I. Rikos$^{2}$\orcidlink{0000-0002-8737-1984}, Ning Pang$^{3}$\orcidlink{0000-0003-4322-4314}, and Yiding Ji$^{1*}$\orcidlink{0000-0003-2678-7051}
		\thanks{$^{1}$Z. Wang, Y. Gao, and Y. Ji are with Robotics and Autonomous Systems Thrust, The Hong Kong University of Science and Technology (Guangzhou), Guangzhou, China. (E-mails: zwang216@connect.hkust-gz.edu.cn, y.gao@gaoyunailab.com,
jiyiding@hkust-gz.edu.cn).}
		\thanks{$^{2}$A. I. Rikos is with the Artificial Intelligence Thrust, The Hong Kong University of Science and Technology (Guangzhou), Guangzhou, China. (Email: apostolosr@hkust-gz.edu.cn)}
		\thanks{$^{3}$N. Pang is with the Department of Automation, Shanghai Jiao Tong University, Shanghai, China. (Email: ningpangswu76@163.com)}
	\thanks{National Natural Science Foundation of China supports this work grants 62303389 and 62373289; Guangdong Basic and Applied Research Funding grants 2022A151511076 and 2024A1515012586; Guangdong Research Platform and Project Scheme grant 2024KTSCX039; Guangzhou Basic and Applied Basic Research Scheme grant 2023A04J1067; Guangzhou-HKUST(GZ) Joint Funding grants 2023A03J0678, 2023A03J0011, 2024A03J0618, 2024A03J0680 and 2025A03J3960.}
    }

	\maketitle
	
	\begin{abstract}
		This paper investigates event-triggered consensus tracking in nonlinear semi-strict-feedback multi-agent systems involving one leader and multiple followers. We first employ radial basis function neural networks and backstepping techniques to approximate the unknown nonlinear dynamics, facilitating the design of dual observers to measure the unknown states and disturbances. Then three adaptive event-triggered control schemes are proposed: fixed-threshold, relative-threshold, and switched-threshold configurations, each featuring specialized controller architectures and triggering mechanisms. Through Lyapunov stability analysis, we establish that the follower agents can asymptotically track the reference trajectory of the leader, meanwhile all error signals remain uniform bounded. Our proposed control strategies effectively prevent Zeno behaviors through stringent exclusion criteria. Finally, an illustrative example is presented, demonstrating the competitive performance of our control framework in achieving consensus tracking and optimizing triggering efficiency.

	\end{abstract}
	
    
	\vspace{-10pt}
	\section{Introduction}
	For decades, consensus tracking in nonlinear multi-agent systems (MASs) has been intensively studied in control and learning communities due to its wide applications such as power grids \cite{intro1,intro2}, intelligent transportation \cite{intro3,intro3.5,intro4} and communication systems \cite{intro5,intro6}. The primary objective is to ensure that all units in the network operate ``unanimously" along a desired trajectory and achieve consensus~\cite{intro7}. However, the presence of uncertain system parameters and external perturbations in real engineering scenarios poses considerable challenges for consensus tracking, often rendering existing methods fail to work. Consequently, it is critical to develop control methods that effectively address the issues.\par

	In the conventional settings of the sample-data control system, controllers continuously respond to changes in the system, which often results in unnecessary resource usage, particularly when the network has restricted bandwidth. A seminal work~\cite{eve1} demonstrated the advantages of event-triggered control over periodic impulse control in first-order stochastic systems with multiplicative noise, leading to its widespread adoption in both linear and nonlinear systems. Recent advancements in controller design such as \cite{eve4,eve5,eve1,eve2,eve3,eve3.1,eve3.2,eve3.3,pang,chufa,jiang2024} have further propelled the field of event-triggered control.\par


   The application of event-triggered control to nonlinear systems has achieved notable success in reducing trigger frequency and enhancing control performance \cite{eve4,eve5}. This framework has since been extended by introducing new triggering conditions. For instance, \cite{eve4} proposed a dynamic triggering condition that alleviates the traditional periodic execution requirements in closed-loop nonlinear systems.
   The work in \cite{eve5} utilized the current state of the plant to determine triggering time instances without the requirement of periodicity. More recently, event-triggered mechanisms have been extensively explored within the context of MASs. Notably, \cite{eve2} introduced a saturated-threshold event-triggered control strategy which aims to achieve robust consensus tracking in continuous-time nonlinear MASs amidst sensor attacks, and \cite{eve3} employed a switching-based trigger strategy for consensus tracking. Additionally, \cite{chufa} analyzed three event-triggered control strategies in nonlinear uncertain systems, which did not require input-to-state stability (ISS).

    In light of these results, we propose an adaptive consensus tracking framework for high-order nonlinear MASs, which integrates the backstepping method, filtering techniques, radial basis function neural networks (RBF NNs), and observer design. 
    Our key contributions are summarized as follows:

	
	\begin{itemize}[leftmargin=*]
		\item We employ RBF NNs with backstepping method to approximate the unknown transition functions of the MASs;
        
		\item We develop observers to monitor both unknown states and unmeasured disturbances, which later enhance the robustness of the controller against perturbations;
        
		\item We propose three event-triggered control strategies: fixed-threshold, relative-threshold, and switched-threshold, then highlight their performance in terms of reduced control update frequency, improved resource utilization, and extended controller lifespan in the empirical study.
	\end{itemize}
    
    The remainder of the work is structured as follows. Section~\ref{sec:preliminary} reviews preliminary knowledge of MAS, observer design and graph theory, then formulates the key problem of the work. Section~\ref{sec:method} develops the fixed-relative-switched threshold control strategies and analyzes the stability of the strategies. Section~\ref{sec:example} presents simulation results to validate the performance of our method. Finally, Section~\ref{sec:conclusion} concludes the paper and lists several future research directions.
	
	\section{Preliminaries and Problem Formulation}~\label{sec:preliminary}
    \vspace{-20pt}
    
    \subsection{System Model}
    
	Consider a team of agents with $N$ followers indexed from $1$ to $N$ and a leader indexed $0$, they communicate with each other in a directed graph and form the MAS model~\cite{tnnls} \cite{mu}:

    \vspace{-10pt}
	\begin{align}\label{equ:mas}
		\dot{x}_{i,k}&=x_{i,k+1}+f_{i,k}(\bar{x}_{i,k})+\xi_{i,k} \nonumber\\
		\dot{x}_{i,n}&=u_i+f_{i,n}(\bar{x}_{i,n})+\xi_{i,n} \nonumber\\
		y_i&=x_{i,1}
	\end{align}
	where $i=1,...,N$, $j=1,...,n$, $k=1,...,n-1$,  $\bar{x}_{i,j}=[x_{i,1},...,x_{i,j}]^T$ are the state of the $i$th follower, while $u_i\in R$ denotes its input of control. The output of the $i$th follower is expressed as $y_i \in R$. The function $f_{i,j}(\bar{x}_{i,j})$ represents $C^1$ class nonlinear smooth equation vectors. $\xi_{i,j}$ represents unmeasured external perturbations affecting the system. Notably, the leader's movement occurs independently without being influenced by the actions or positions of the followers.
	
	\subsection{Observer Design}
	To effectively locate the unknown states and perturbations throughout the entire control system, we introduce a series of observers. By precisely analyzing and processing feedback data, these observers accurately identify and estimate unknown variables within the system, enhancing the overall robustness and stability. The state observer is defined as
    
    \vspace{-10pt}
	\begin{align}
		\dot{\hat{\bar{x}}}_{i,n}&=(P_i\otimes I_m)\dot{\hat{\bar{x}}}_{i,n}+(Q_i\otimes y_i)+\sum_{l=1}^{n}(R_{i,l}\otimes\hat{f}_{i,l}(\dot{\hat{\bar{x}}}_{i,l}))\nonumber\\
		&+(U_i\otimes u_i)+\hat{\varpi}_i  \nonumber\\
		y_i&=(V_i^T\otimes I_m)\hat{\bar{x}}_n
	\end{align}
    where $\otimes$ represents the Kronecker product, and ${{\hat{\bar{x}}}_{i,l}}={{[ \hat{x}_{i,1}^{T },\ldots ,\hat{x}_{i,l}^{T } ]}^{T }}$ represents the estimated value of the actual state with $l=1,...,n$. The vector $Q_i=[q_{i,1},...,q_{i,n}]^T$ is such that the matrix $P_i$ is a strictly Hurwitz. The parameters are defined as follows: $R_{i,l}=[0,\ldots ,1,\ldots ,0 ]_{n\times 1}^T$ where the $l$-th element is $1$, $U_i=[0,...,0,1]^T_{n\times 1}$, $V_i=[1,0,...,0]^T_{n\times1}$ and $P_i=[-Q_i, R_{i, 1},\ldots, R_{i,n-1}]$.

	We denote by $\psi_i=[(x_{i,1}-\hat{x}_{i,1})^T,...,(x_{i,n}-\hat{x}_{i,n})^T]^T$ the error of state observation and have that
	\begin{align}
		\dot{\psi}_i=(P_i\otimes I_m)\psi_i+\widetilde{\varpi}_i+\sum_{l=1}^{n}S_{i,l}\otimes(f_{i,l}(\overline{x}_{i,l})-\hat{f}_{i,l}(\hat{\overline{x}}_{i,l}))
	\end{align}
	
	We also denote the error of function approximation by ${{\varphi}_{i}}(t)=\sum\limits_{l=1}^{n}{{{R}_{i,l}}\otimes ( {{f}_{i,l}}\left( {{{\bar{x}}}_{i,l}}\left( t \right) \right)-{{{\hat{f}}}_{i,l}}( {{{\hat{\bar{x}}}}_{i,l}}( t ) ) )}$ and write ${{\varphi }_{i}}(t)={{\left[ \varphi_{i,1}^{T }(t),...,\varphi _{i,n}^{T }(t) \right]}^{T }}$. Suppose that ${{\varphi }_{i}}(t)$ is bounded, thus there exists an unspecified parameter $\varphi _i^0>0$ such that the inequality $\left\| {{\varphi_i}(t)} \right\| \le \varphi_i^0$ holds. Using RBF NNs to approximate the unidentified nonlinear function of the MASs~(\ref{equ:mas}), an optimal weight vector $W_{i,l}^{*}$ is derived as:
	\begin{align}
		{{f}_{i,l}}({{\bar{x}}_{i,l}})=W_{i,l}^{*T }{{E}_{i,l}}\left( {{{\hat{\bar{x}}}}_{i,l}} \right)+{{\sigma}_{i,l}}(t)\label{w1}
	\end{align}
	where ${{\sigma}_{i,l}}(t)$ represents the bounded approximation error, i.e., there is a parameter ${{\sigma}_{0,l}}>0$ such that $\left| {{\sigma}_{i,l}}(t) \right|\le {{\sigma}_{0,l}}$.	
    
    Let $\hat{W}_{i,l}^{T}$ be the estimation of $W_{i,l}^{*T }$, then the estimated smooth nonlinear function is ${{\hat{f}}_{i,l}}({{\hat{\bar{x}}}_{i,l}})=\hat{W}_{i,l}^{T }{E_{i,l}}\left( {{{\hat{\bar{x}}}}_{i,l}}\right)$. The optimal weight $W_{i,l}^{*}$ for $l=1,\ldots ,n$ is designed as:

    \vspace{-10pt}
	\begin{equation}
		W_{i,l}^{*}=\underset{{}}{\mathop{\arg }}\,\underset{\hat{W}_{i,l}^{{}}\in \overset{\lower0.5em\hbox{$\smash{\scriptscriptstyle\smile}$}}{\Omega }}{\mathop{\min }}\, \underset{{}}{\mathop{\underset{\begin{smallmatrix}
						{{{\bar{x}}}_{i,l}}\in {{\Omega }_{i,l}} \\
						{{{\hat{\bar{x}}}}_{i,l}}\in {{{\hat{\Omega }}}_{i,l}}
				\end{smallmatrix}}{\mathop{\sup }}\,\left| {{{\hat{f}}}_{i,l}}({{{\hat{\bar{x}}}}_{i,l}})-f({{{\bar{x}}}_{i,l}}) \right|}} \label{w2}
        \vspace{-20pt}
	\end{equation}
	where ${{\Omega }_{i,l}}$, ${{{\hat{\Omega }}}_{i,l}}$ and ${\overset{\lower0.5em\hbox{$\smash{\scriptscriptstyle\smile}$}}{\Omega }}$ represent compact sets corresponding to ${{\bar{x}}_{i,l}}$, ${{\hat{\bar{x}}}_{i,l}}$, $\hat{E}_{i,l}^{{}}$.
	Besides, define $\Theta _{i}^{*}=\max \left\{ \left\| W_{i,l}^{*} \right\| \right\}$, where  ${{\hat{\Theta }}_{i}}$ represents the estimation of $\Theta _{i}^{*}$ with ${{\tilde{\Theta }}_{i}}=\Theta _{i}^{*}-{{\hat{\Theta }}_{i}}$.
	
	The state observers is then revised as: $\forall 1\leq k<n$
	\begin{align}
		& {{{\dot{\hat{x}}}}_{i,k}}={{{\hat{x}}}_{i,k+1}}+\hat{W}_{i,k}^{T }{{E}_{i,k}}\left( {{{\hat{\bar{x}}}}_{i,k}} \right)+{{q}_{i,k}}{{\psi}_{i,1}}+{{{\hat{\varpi}}}_{i,k}} \nonumber\\
		& {{{\dot{\hat{x}}}}_{i,n}}={{u}_{i}}+\hat{W}_{i,n}^{T }{{E}_{i,n}}\left( {{{\hat{\bar{x}}}}_{i,n}} \right)+{{q}_{i,n}}{{\psi}_{i,1}}+{{{\hat{\varpi }}}_{i,n}}
	\end{align}
	
	Next, we define the estimate of unknown external perturbations as $\hat{\varpi}_i=[\hat{\varpi}_{i,1}^T,...,\hat{\varpi}_{i,n}^T]^T$. In order to achieve consensus control, we introduce an auxiliary variable ${\tau}_{i,l}={\varpi}_{i,l}-{\kappa}_{i,l}{x}_{i,l}$ where $\kappa_{i,l}$ is a positive parameter. 
	
	Given the above concepts, the perturbation observer is:
    \vspace{-5pt}
	\begin{align}
		{{{\hat{\varpi}}}_{i,l}}&={{{\hat{\tau}}}_{i,l}}+{{\kappa }_{i,l}}{{{\hat{x}}}_{i,l}} \nonumber\\
		{{{\dot{\hat{\tau}}}}_{i,l}}&=-{{\kappa}_{i,l}}( \hat{W}_{i,l}^{T }{{E}_{i,l}}({{{\hat{\bar{x}}}}_{i,l}})+{{{\hat{\tau}}}_{i,l}}+{{\kappa}_{i,l}}{{{\hat{x}}}_{i,l}}+{{{\hat{x}}}_{i,l+1}})
	\end{align}
	where ${{x}_{i,n+1}}={{u}_{i}}$ and $\hat W_{i,l}^{}$ is the estimation of $W_{i,l}^*$.
	
	\subsection{Directed Graph}
    Given MASs with $N$ followers and one leader, the interaction between agents is modeled by a directed graph $\mathcal{G}=(\mathcal{V},\mathcal{E})$, where $\mathcal{V}=\{1,...,N\}$ represents the set of nodes, $\mathcal{E}\subseteq \mathcal{V}\times\mathcal{V}$ represents the set of edges, and $(\mathcal{V}_i,\mathcal{V}_j)\in \mathcal{E}$ indicates that agent $j$ can gain information from agent $i$. The adjacent matrix $\mathcal{A}=[a_{i,j}]\in R^{N\times N}$ of $\mathcal G$ is defined as:
	\begin{equation}
		a_{i,j}=\left\{
		\begin{aligned}
			&1, \text{ } if \text{ }(\mathcal{V}_i, \mathcal{V}_j)\in \mathcal{E}\\
			&0, \text{ } if \text{ }(\mathcal{V}_i, \mathcal{V}_j)\notin \mathcal{E}
		\end{aligned}
		\right.
	\end{equation}

     We introduce the Laplacian matrix $\mathcal{L}$ as $\mathcal{L}=\mathcal{D}-\mathcal{A}$, where $\mathcal{D}=diag( {{d_1},...,{d_N}})$ is the in-degree matrix of graph $\mathcal{G}$, ${d_i} = \sum\nolimits_{j = 1,j \ne i}^N {{a_{ij}}}$. And clarify the diagonal matrix $\mathcal{B}  = diag\left( {{\mathcal{C}_1},...,{\mathcal{C}_N}} \right)$, if the leader 0 is capable of transmitting information to agent $i$, then ${\mathcal{C}_i} > 0$. If not, ${\mathcal{C}_i} = 0$ holds.

    \subsection{Problem Formulation}
    \vspace{-10pt}
     \begin{figure}[ht!]
		\centering
		\includegraphics[height=3cm,width=8.5cm]{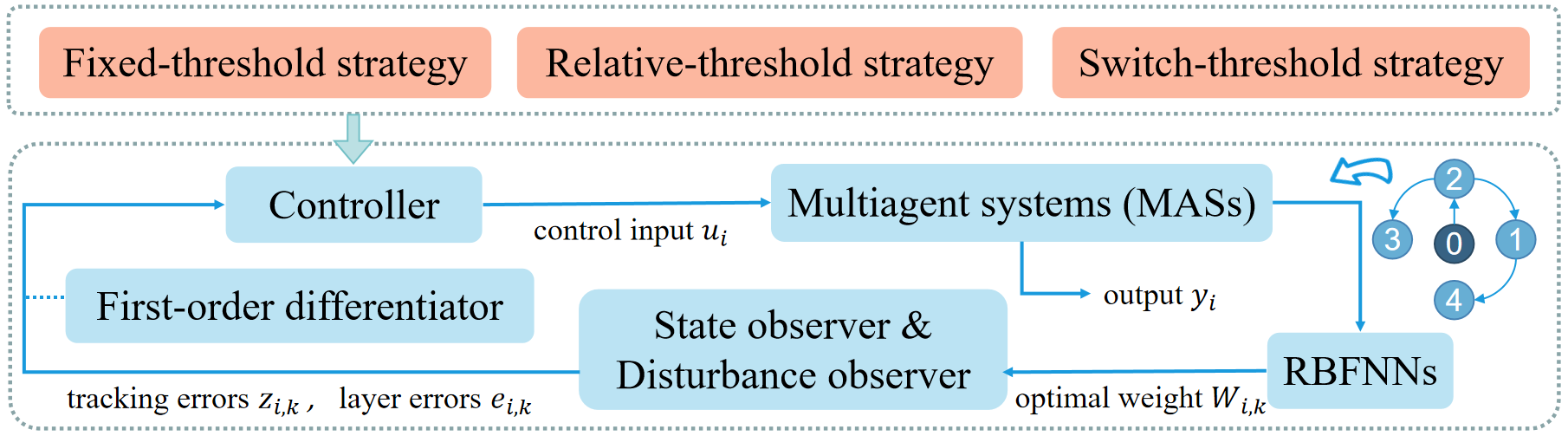}
        \vspace{-10pt}
		\caption{Block diagram of our proposed control framework.}
	\end{figure}
    \vspace{-10pt}

 Fig. 1 illustrates our framework of adaptive event-triggered consensus tracking control. The following lemmas and assumptions are necessary to facilitate the later discussions.

	\begin{lemma}[from~\cite{lemma1}]
	For any positive definite matrix ${{H}_{i}}=H_{i}^{T }>0$, with a symmetric positive matrix ${{F}_{i}}$ and a strict Hurwitz matrix $P_i$, it satisfies $P_{i}^{T }{{F}_{i}}+{{F}_{i}}{{P}_{i}}=-2{{H}_{i}}$.
	\end{lemma}

    \begin{lemma}[from~\cite{lemma2}]
     Inequality $0\le \left| \chi  \right|-\chi \tanh ( \frac{\chi }{{{\chi}_{0}}} )\le 0.2785{{\chi}_{0}}$
	holds for any parameter $\chi \in R$ and ${{\chi}_{0}} >0$.
    \end{lemma}

    \vspace{-8pt}
	\begin{assumption}[from~\cite{eve3.2, tnnls, rela2}]
	The unmeasured external perturbations are all bounded, i.e., inequality $\left\| {{\varpi}_{i,j}} \right\|\le \varpi _{i,j}^{0}$ holds, where $\varpi_{i,j}^{0}$ is a positive parameter.  
	\end{assumption}

    \vspace{-8pt}
	\begin{assumption}[from~\cite{mu}]
	    The desired reference signal of the leader in the MASs is both measurable and smooth, also ${{y}_{r}}\left( t \right)$ and its first order derivative ${{\dot{y}}_{r}}\left( t \right)$ are bounded.
	\end{assumption}
	
    \vspace{-8pt}
    \begin{problem}[Consensus tracking of MASs]\label{prob:consensus}
      Given a class of nonlinear MASs following equation~(\ref{equ:mas}), suppose that the communication of agents is modeled by a directed graph $\mathcal G$, each follower is subject to unknown external perturbations $\xi$ and required to track the leader's reference signal $y_r$, then our goals are two-fold: (i) design an adaptive control law to ensure that the tracking errors are uniformly bounded; (ii) evaluate the performance of different event-triggered threshold control strategies in terms of control update frequency, resources conservation and controller's lifespan.
    \end{problem}

	\section{Event-Triggered Controller Design}\label{sec:method}
    
	This section is initiated by the general procedure of controller design for Problem~\ref{prob:consensus}. First, the graph-based errors $z_{i,k}$ and boundary layer errors $e_{i,k}$ for the $i$-th follower are:
    \vspace{-10pt}
	\begin{align}
		z_{i,1}&=\sum_{j=1}^{N}a_{i,j}(y_i-y_j)+b_i(y_i-y_r) \nonumber\\
		z_{i,k}&=x_{i,k}-\bar{\alpha}_{i,k} \nonumber\\
		e_{i,k}&=\bar{\alpha}_{i,k}-\alpha_{i,k}
	\end{align}
	where $k=2,...,n$. The terms $\alpha_{i,k}$ and $\bar{\alpha}_{i,k}$ respectively denote the
	virtual control and its filtered counterpart.
	
	$Step\text{ }1$: The Lyapunov function is determined as
	
	\vspace{-10pt}
	\begin{align}
		{{V}_{i,1}}&=\frac{z_{i,1}^{2}}{2}+\frac{\tilde{\tau}_{i,1}^{2}}{2}+\frac{1}{2{{\eta }_{i,1}}}\tilde{W}_{i,1}^{T }{{\tilde{W}}_{i,1}}+\frac{1}{2{{o}_{i}}}\tilde{\Theta }_{i}^{2}+{{V}_{0}}\nonumber\\
		V_0&=\frac{1}{2}\psi_i^T(F_i\otimes I_m)\psi_i
	\end{align}
	where $\psi_i=[\psi_{i,1}^T,...,\psi_{i,n}^T]^T$, $\eta_{i,1}>0$ and $o_i>0$ define designed parameters. Then, define an auxiliary function as

     \vspace{-10pt}
	\begin{align}
		{{\bar{Z}}_{i,1}}\left( {{T}_{i,1}} \right)=&-\sum\limits_{j=1}^{N}{{{a}_{i,j}}}\left( {{{\hat{x}}}_{j,2}}+{{f}_{j,1}}\left( {{{\bar{x}}}_{j,1}} \right)+{{\varpi}_{j,1}}+{{\psi}_{j,2}} \right)\nonumber\\&+\left( {{d}_{i}}+{{b}_{i}} \right)\left( {{f}_{i,1}}\left( {{{\bar{x}}}_{i,1}} \right)+{{\varpi }_{i,1}}+{{\psi}_{i,2}} \right)-{{b}_{i}}{{\dot{y}}_{r}}.
	\end{align}
	
	${{\bar{Z}}_{i,1}}\left( {{T}_{i,1}} \right)$ is estimated by RBF NNs: ${{\bar{Z}}_{i,1}}\left( {{T}_{i,1}} \right)=K_{i,1}^{*T }{{E}_{i,1}}\left( {{T}_{i,1}} \right)+{{\delta}_{i,1}}\left( {{T}_{i,1}} \right)$, where $\bar{\delta}_{i,1}$ defines an unknown positive parameter, ${{T}_{i,1}}={{[{{y}_{r}},{{\dot{y}}_{r}},\hat{x}_{i,1}^{T},\hat{x}_{j,1}^{T},{{\hat{\varpi}}_{i,1}}^{T},{{\hat{\varpi }}_{j,1}}^{T}]}^{T}}$ denotes the input of RBF NNs and $K_{i,1}^{*T }$ is the optimal weight, just like the $W_{i,1}^{*T}$ in (\ref{w1})(\ref{w2}), with $\left| {{\delta}_{i,1}}\left( {{T}_{i,1}} \right) \right|\le\bar{\delta}_{i,1}$.
	
	Based on the Young's inequality, with a positive designed parameter ${c}_{i,1}$, the subsequent inequality can be derived:
     
     \vspace{-5pt}
	\begin{align}
		{{z}_{i,1}}{{\bar{Z}}_{i,1}}\left( {{T}_{i,1}} \right)\le& \frac{\Theta _{i}^{*}}{2c_{i,1}^{2}}z_{i,1}^{2}E_{i,1}^{T }\left( {{T}_{i,1}} \right){{E}_{i,1}}\left( {{T}_{i,1}} \right)\nonumber\\&+\frac{c_{i,1}^{2}}{2}+\frac{z_{i,1}^{2}}{2}+\frac{\bar{\delta}_{i,1}^{2}}{2}
	\end{align}
	
	By Young's inequality and Assumption 1, one has
     \vspace{-5pt}
	\begin{align}
		\psi_{i}^{T }\left( {{F}_{i}}\otimes {{I}_{m}} \right){{\varphi }_{i}}\le& \frac{1}{2}{{\left\| {{F}_{i}}\otimes {{I}_{m}} \right\|}^{2}}{{\left\| \varphi_{i}^{0} \right\|}^{2}}+\frac{1}{2}{{\left\| {{\psi}_{i}} \right\|}^{2}}\\
		\psi_{i}^{T }\left( {{F}_{i}}\otimes {{I}_{m}} \right){{\tilde{\varpi}}_{i}}\le& \frac{1}{2}{{\left\| {{F}_{i}}\otimes {{I}_{m}} \right\|}^{2}}{{\left\| \tilde{\varpi}_{i}^{0} \right\|}^{2}}+\frac{1}{2}{{\left\| {{\psi }_{i}} \right\|}^{2}}
	\end{align}
	
	The virtual and adaptive control laws are formulated as
     \vspace{-5pt}
	\begin{align}
		{{\alpha}_{i,2}}=&\frac{1}{{{d}_{i}}+{{b}_{i}}}( {{r}_{i,1}}{{z}_{i,1}}-\frac{{{z}_{i,1}}}{2}\nonumber \\
		&-\frac{{{{\hat{\Theta }}}_{i}}}{2c_{i,1}^{2}}z_{i,1}E_{i,1}^{T}\left( {{T}_{i,1}} \right){{E}_{i,1}}\left( {{T}_{i,1}} \right))\\
		{{\dot{\hat{W}}}_{i,1}}=&-{{h}_{i,1}}{{\hat{W}}_{i,1}}-{{\eta }_{i,1}}{{\tilde{\tau}}_{i,1}}{{\kappa }_{i,1}}{{E}_{i,1}}\left( {{{\hat{\bar{x}}}}_{i,1}} \right)
	\end{align}
	where ${{h}_{i,1}}$ and ${{r}_{i,1}}$ are designed parameters. Then, considering Lemma 1 and the above formulas, one has
    \vspace{-5pt}
	\begin{align}
		\dot{V}_{i,1}\le& {{r}_{i,1}}z_{i,1}^{2}+\left( {{d}_{i}}+{{b}_{i}} \right){{z}_{i,1}}\left( {{z}_{i,2}}+{{e}_{i,2}} \right)+{{\tilde{\tau}}_{i,1}}{{\dot{\varpi }}_{i,1}}\nonumber\\&+( \frac{3-2{{\kappa}_{i,1}}}{2} )\tilde{\tau}_{i,1}^{2}+{{\iota }_{i,1}}+\frac{{{h}_{i,1}}}{{{\eta }_{i,1}}}\tilde{W}_{i,1}^{T }{{\hat{W}}_{i,1}}\nonumber\\&+{{\tilde{\Theta }}_{i}}( \frac{1}{2c_{i,1}^{2}}z_{i,1}^{2}E_{i,1}^{T}( {{T}_{i,1}} ){{E}_{i,1}}( {{T}_{i,1}} )-\frac{{{{\dot{\hat{\Theta }}}}_{i}}}{{{o}_{i}}} )\nonumber\\&-\psi_{i}^{T }( {{H}_{i}}\otimes {{I}_{m}} ){{\psi}_{i}}+( 1+\kappa_{i,1}^{2}+\kappa _{i,1}^{4} ){{\left\| {{\psi}_{i}} \right\|}^{2}}
	\end{align}
	where
	${{\iota }_{i,1}}=\frac{c_{i,1}^{2}}{2}+\frac{\bar{\delta} _{i,1}^{2}}{2}+\frac{\kappa_{i,1}^{2}\sigma _{0,1}^{2}}{2}+\frac{1}{2}{{\left\| \varphi_{i}^{0} \right\|}^{2}}{{\left\| {{F}_{i}}\otimes {{I}_{m}} \right\|}^{2}}+\frac{1}{2}{{\left\| \tilde{\varpi}_{i}^{0} \right\|}^{2}}{{\left\| {{F}_{i}}\otimes {{I}_{m}} \right\|}^{2}}$. Then, with a small positive designed parameter $m_{i,2}$, process ${{\alpha}_{i,2}}$ through the first-order low-pass filter, resulting in ${{\bar{\alpha}}_{i,2}}$ as follows:
    \vspace{-5pt}
	\begin{align}
		&{{\bar{\alpha}}_{i,2}}\left( 0 \right)={{\alpha}_{i,2}}\left( 0 \right)\nonumber\\
		&{{m}_{i,2}}{{\dot{\bar{\alpha}}}_{i,2}}+{{\bar{\alpha}}_{i,2}}={{\alpha}_{i,2}}
	\end{align}
	
	 $Step\text{ }k$: Choose the Lyapunov function as
	
	\vspace{-10pt}
	\begin{align}
		{{V}_{i,k}}=\frac{z_{i,k}^{2}}{2}+\frac{\tilde{\tau}_{i,k}^{2}}{2}+\frac{1}{2{{\eta }_{i,k}}}\tilde{W}_{i,k}^{T }{{\tilde{W}}_{i,k}}
	\end{align}
	where ${{\eta }_{i,k}}>0$ is a parameter. By utilizing RBF NNs, one has ${{\bar{Z}}_{i,k}}\left( {{T}_{i,k}} \right)=K_{i,k}^{*T }{{E}_{i,k}}\left( {{T}_{i,k}} \right)+{{\delta}_{i,k}}\left( {{T}_{i,k}} \right)$, where $\bar{\delta}_{i,k}>0$ is an unknown parameter, with $|{\delta}_{i,k}(T_{i,k})|\le\bar{\delta}_{i,k}$. Then we design the virtual control law and adaptive law as:
    \vspace{-5pt}
	\begin{align}
		{\alpha_{i,k + 1}} =&  {r_{i,k}}{z_{i,k}} - \frac{{{z_{i,k}}}}{2}+\frac{{{\alpha_{i,k}} - {{\bar \alpha}_{i,k}}}}{{{m_{i,k}}}}- {q_{i,k}}{\psi_{i,1}} \nonumber\\&- \frac{{{{\hat{\Theta }}}_{i}}}{{2c_{i,k}^2}}{z_{i,k}}E_{i,k}^ T \left( {{T_{i,k}}} \right){E_{i,k}}\left( {{T_{i,k}}} \right) \\
		{{\dot{\hat{W}}}_{i,k}}=&-{{h}_{i,k}}{{\hat{W}}_{i,k}}-{{\eta }_{i,k}}{{\tilde{\tau}}_{i,k}}{{\kappa}_{i,k}}{{E}_{i,k}}\left( {{{\hat{\bar{x}}}}_{i,k}} \right)
	\end{align}
	where ${{T}_{i,k}}={{[{{y}_{r}},{{\hat{\Theta }}_{i}},\hat{\bar{x}}_{i,k}^{T},\hat{\bar{x}}_{j,k}^{T},\hat{\varpi }_{i,k}^{T},\hat{\varpi}_{j,k}^{T}]}^{T}}$, ${{r}_{i,k}}$ and ${{h}_{i,k}}$ are designed parameters. By the Young's inequality, one has
    \vspace{-10pt}
	\begin{align}
		{{\dot{V}}_{i,k}}\le&{{r}_{i,k}}z_{i,k}^{2}+{{\left(\kappa _{i,k}^{2}+\kappa_{i,k}^{4} \right)\left\| {{\psi}_{i}} \right\|}^{2}}+ {{z}_{i,k}}\left( {{z}_{i,k+1}}+{{e}_{i,k+1}} \right)\nonumber\\&+\frac{{{h}_{i,k}}}{{{\eta }_{i,k}}}\tilde{W}_{i,k}^{T }{{\hat{W}}_{i,k}}+( \frac{3-2{{\kappa}_{i,k}}}{2} )\tilde{\tau}_{i,k}^{2}+{{\tilde{\tau}}_{i,k}}{{\dot{\varpi }}_{i,k}}\nonumber\\&+\frac{{{{\tilde{\Theta }}}_{i}}}{2c_{i,k}^{2}}z_{i,k}^{2}E_{i,k}^{T }\left( {{T}_{i,k}} \right){{E}_{i,k}}\left( {{T}_{i,k}} \right)+{{\iota }_{i,k}}
	\end{align}
	where
	${{\iota }_{i,k}}=\frac{c_{i,k}^{2}}{2}+\frac{\bar{\delta} _{i,k}^{2}}{2}+\frac{\kappa_{i,k}^{2}\sigma_{0,k}^{2}}{2}$ and $c_{i,k}0>0$. In line with Step\text{ }1, the first-order low-pass filter is implemented as:
    \vspace{-5pt}
	\begin{align}
		&{{\bar{\alpha}}_{i,k+1}}\left( 0 \right)={{\alpha}_{i,k+1}}\left( 0 \right)\nonumber\\&{{m }_{i,k+1}}{{\dot{\bar{\alpha}}}_{i,k+1}}+{{\bar{\alpha}}_{i,k+1}}={{\alpha}_{i,k+1}}
	\end{align}
	where  ${{m}_{i,k+1}}$ denotes a slight positive designed parameter.
	
	$Step\text{ }n$: Choose the Lyapunov function as
	
	\vspace{-10pt}
	\begin{align}
		{{V}_{i,n}}=\frac{z_{i,n}^{2}}{2}+\frac{\tilde{\tau}_{i,n}^{2}}{2}+\frac{1}{2{{\eta }_{i,n}}}\tilde{W}_{i,n}^{T }{{\tilde{W}}_{i,n}}
	\end{align}
	where ${{\eta }_{i,n}}$ denotes a designed positive parameter. By RBF NNs, one has ${{\bar{Z}}_{i,n}}\left( {{T}_{i,n}} \right)=K_{i,n}^{*T }{{E}_{i,n}}\left( {{T}_{i,n}} \right)+{{\delta}_{i,n}}\left( {{T}_{i,n}} \right)$, where $\bar{\delta}_{i,n}$ denotes an unknown positive parameter, with $\left| {{\delta}_{i,n}}\left( {{T}_{i,n}} \right) \right|\le\bar{\delta}_{i,n}$. With designed parameters ${{r}_{i,n}}$, ${{h}_{i,n}}$, ${{\lambda}}_{i}$, the control law and adaptive parameters are designed as
    \vspace{-5pt}
	\begin{align}
		&{\alpha_{i,n+1}} =r_{i,n}{z_{i,n}}-\frac{{{z_{i,n}}}}{2}+\frac{{{\alpha_{i,n}} - {{\bar \alpha}_{i,n}}}}{{{m_{i,n}}}}- {q_{i,n}}{\psi_{i,1}} \nonumber\\&- \frac{{{{\hat{\Theta }}}_{i}}}{{2c_{i,n}^2}}{z_{i,n}}E_{i,n}^ T \left( {{T_{i,n}}} \right){E_{i,n}}\left( {{T_{i,n}}} \right)\label{a1} \\
		&{{\dot{\hat{W}}}_{i,n}}=-{{h}_{i,n}}{{\hat{W}}_{i,n}}-{{\eta }_{i,n}}{{\tilde{\tau}}_{i,n}}{{\kappa}_{i,n}}{{E}_{i,n}}\left( {{{\hat{\bar{x}}}}_{i,n}} \right)\label{a2}\\
		&{{\dot{\hat{\Theta }}}_{i}}=-{{{\lambda}}_{i}}{{\hat{\Theta }}_{i}}+\sum\limits_{k=1}^{n}{\frac{{{o }_{i}}}{2c_{i,k}^{2}}z_{i,k}^{2}E_{i,k}^{T }\left( {{T}_{i,k}} \right){{E}_{i,k}}\left( {{T}_{i,k}} \right)}\label{a3}
	\end{align}
    
We now develop three event-triggered control strategies that provably ensure consensus tracking for the MASs. Theoretical analysis is also provided for their convergence.

	\subsection{Fixed-threshold strategy}
	The adaptive event-triggered controller is reformulated as	
    \vspace{-10pt}
	\begin{align}
		{{w}_{i}}\left( t \right)={{\alpha}_{i,n+1}}-{{\bar{\pi}}_{i}}\tanh ( \frac{{{z}_{i,n}}{{{\bar{\pi}}}_{i}}}{{{\mu}_{i}}} )\label{fixed0}
	\end{align}
where the triggering condition is defined as
\vspace{-5pt}
	\begin{align}
		{{u}_{i}}\left( t \right)=&w_{i}\left( {{t}_{s}} \right)\text{ },\text{  }\forall \text{t}\in \left[ {{t}_{s}},{{t}_{s+1}} \right)\label{fixed1}\\
		{{t}_{s+1}}=&\inf \left\{ \text{t}\in R||{{\vartheta}_{i}}\left( t \right)|\ge {{\pi}_{i}} \right\}\text{,  }{{\text{t}}_{1}}=0\label{fixed2}
	\end{align}
	where $\vartheta_{i}(t)=w_i(t)-u_i(t)$ is the measurement-triggered error and $t_s$ is the controller's update time with $s\in Z$. The parameters $\bar{\pi}_i$, $\pi_i$ and $\mu_i$ are designed to be positive with $\bar{\pi}>\pi_i$. The control signal ${{u}_{i}}\left( {{t}_{s+1}} \right)$ is applied to the system when condition (\ref{fixed2}) is activated. During the interval $\text{t}\in \left[ {{t}_{s}},{{t}_{s+1}} \right)$, i.e. $\left| {{w }_{i}}\left( t \right)-{{u}_{i}}\left( t \right) \right|< {{\pi}_{i}}$, the controller maintains a constant value of ${{w}_{i}}\left( {{t}_{s}} \right)$. For a function ${{\epsilon}_{i}} {{\left(t\right)}}$ that continuously changes over time and satisfies ${{\epsilon}_{i}}\left( {{t}_{s}} \right)=0$ and ${{\epsilon }_{i}}\left( {{t}_{s+1}} \right)=\pm 1$ with $\left| {{\epsilon}_{i}}\left( t \right) \right|\le 1$, it follows that ${{w}_{i}}\left( t \right)={{u}_{i}}\left( t \right)+{{\epsilon}_{i}}\left( t \right){{\pi}_{i}}$. By Lemma 2, $-{{\epsilon }_{i}}(t){{\pi}_{i}}{{z}_{i,n}}-{{\bar{\pi}}_{i}}{{z}_{i,n}}\tanh( \frac{{{z}_{i,n}}{{{\bar{\pi}}}_{i}}}{{{\mu}_{i}}} )\le 0.2875{{\mu}_{i}}$ holds, then we differentiate ${{V}_{i,n}}$ by the Young's inequality and (\ref{a1})-(\ref{a3}), with ${{\iota }_{i,n}}=\frac{c_{i,n}^{2}}{2}+\frac{\bar{\delta} _{i,n}^{2}}{2}+\frac{\kappa_{i,n}^{2}\sigma_{0,n}^{2}}{2}+0.2785{{\mu }_{i}}$ and a positive parameter $c_{i,n}$, and get the result below:
	\begin{align}
		\dot{V}_{i,n}
		\le& {{r}_{i,n}}z_{i,n}^{2}+(\frac{3-2{{\kappa}_{i,n}}}{2})\tilde{\tau}_{i,n}^{2}+{{\tilde{\tau}}_{i,n}}{{\dot{\varpi }}_{i,n}}\nonumber\\&+\frac{{{{\tilde{\Theta }}}_{i}}}{2c_{i,n}^{2}}z_{i,n}^{2}E_{i,n}^{T }\left( {{T}_{i,n}} \right){{E}_{i,n}}\left( {{T}_{i,n}} \right)+{{\iota }_{i,n}}\nonumber\\&+\frac{{{h}_{i,n}}}{{{\eta }_{i,n}}}\tilde{W}_{i,n}^{T }{{\hat{W}}_{i,n}}+{{\left(\kappa _{i,n}^{2}+\kappa_{i,n}^{4} \right)\left\| {{\psi}_{i}} \right\|}^{2}}\label{000}
	\end{align}
	
	\subsection{Relative-threshold strategy}
	In the fixed-threshold strategy, the threshold $\pi_i$ remains constant regardless of the magnitude of the control signal. However, it is preferable to set a variable threshold for the triggering condition to achieve system stabilization~\cite{rela1}. Specifically, when the control signal $u_i$ is huge, a greater measurement error is tolerated, allowing for longer update intervals. Conversely, when $u_i$ approaches $0$, the system states stabilize towards equilibrium and a smaller threshold enables more accurate control, improving the overall performance. Then, we propose the adaptive event-triggered controller:
    
    \vspace{-10pt}
	\begin{align}
		{{w}_{i}}\left( t \right)=&-(1+\Delta_i)({{\alpha}_{i,n+1}}tanh(\frac{z_{i,n}\alpha_{i,n+1}}{\mu_i})\nonumber\\&+{{\bar{\pi}}^*_{i}}\tanh( \frac{{{z}_{i,n}}{{{\bar{\pi}}}^*_{i}}}{{{\mu}_{i}}}))\label{rela0}
	\end{align}
	
	The triggering condition is defined as
    \vspace{-5pt}
	\begin{align}
		{{u}_{i}}\left( t \right)=&w_{i}\left( {{t}_{s}} \right)\text{ },\text{  }\forall \text{t}\in \left[ {{t}_{s}},{{t}_{s+1}} \right)\label{rela1}\\
		{{t}_{s+1}}=&\inf \left\{ \text{t}\in R||{{\vartheta}_{i}}\left( t \right)|\ge \Delta_i|u_i(t)|+{{\pi}^*_{i}} \right\}\label{rela2}
	\end{align}
	where $t_s$ represents the time when the controller is updated, $s\in Z$, $\mu_i$, $\Delta_i$, $0<\Delta_i<1$, $\pi^*_i>0$, and $\bar{\pi}^*_i>\pi^*_i/(1-\Delta_i)$ are all positive designed parameters. From (\ref{rela2}), we have $w_i(t)=(1+\rho_{1,i}(t))u_i(t)+\rho_{2,i}(t)\pi_i^*$ in the interval $[t_s,t_{s+1}]$, where $\rho_{1,i}(t)$ and $\rho_{2,i}(t)$ denote time-varying parameters satisfying $|\rho_{1,i}(t)|\leq1$ and $|\rho_{2,i}(t)|\leq1$. Thus, one has
    \vspace{-10pt}
	\begin{align}
		{{u}_{i}}(t)=\frac{w_i(t)}{1+\rho_{1,i}(t)\Delta_i}-\frac{\rho_{2,i}(t)\pi_i^*}{1+\rho_{1,i}(t)\Delta_i}
	\end{align}
	
	According to the above analysis, similar to (\ref{000}), it can be obtained the same derivative result of $\dot{V}_{i,n}$ with different ${{\iota }_{i,n}}$, where ${{\iota }_{i,n}}=\frac{c_{i,n}^{2}}{2}+\frac{\bar{\delta} _{i,n}^{2}}{2}+\frac{\kappa_{i,n}^{2}\sigma_{0,n}^{2}}{2}+0.557{{\mu }_{i}}$.

	\subsection{Switched-threshold strategy}
	We now introduce a switched-threshold strategy. The relative-threshold approach adjusts the threshold based on the control signal's magnitude, allowing for longer update intervals when the signal is huge and more precise control as the signal nears zero, thus improving performance. However, a very large control signal can lead to significant measurement errors and abrupt changes during updates. In contrast, the fixed-threshold strategy maintains a consistent upper limit on measurement errors, regardless of signal size. Then we propose the control strategy whose switching gate $G$ and the triggering condition are defined below: 
    \vspace{-5pt}
	\begin{align}
		{{u}_{i}}\left( t \right)=w_{i}\left( {{t}_{s}} \right)\text{ },\text{  }\forall \text{t}\in \left[ {{t}_{s}},{{t}_{s+1}} \right)
	\end{align}
	\begin{equation}
		{t}_{s+1}=\left\{
		\begin{aligned}
			&\inf \left\{ \text{t}\in R||{{\vartheta}_{i}}\left( t \right)|\ge \Delta_i|u_i(t)|+{{\pi}^*_{i}} \right\}, |u_i|\geq G\\
			&\inf \left\{ \text{t}\in R||{{\vartheta}_{i}}\left( t \right)|\ge {{\pi}_{i}} \right\}, |u_i|< G
		\end{aligned}
		\right.
	\end{equation}
	
	With $t\in[t_s,t_{s+1})$, we have
    \vspace{-5pt}
	\begin{equation}
		\bar{\vartheta}_i=sup|\vartheta_i(t)|\leq max\{(\Delta_i|u_i(t)|+{{\pi}^*_{i}}),{{\pi}_{i}}\}
	\end{equation}

	Since the switched-threshold strategy employs an identical control law for both the fixed-threshold and relative-threshold strategies, the ultimate bounds for tracking and stabilization errors remain consistent with those of the above two. 
    
    In summary, we have established an adaptive consensus tracking control framework which integrates state-disturbance observer design, learning based dynamics approximation, backstepping method and three event-triggered control strategies. Finally, we prove that all error signals are bounded by employing Lyapunov stability theory. In addition, we also confirm the soundness of the proposed control strategies that effectively avoid Zeno behaviors.

    \begin{theorem}
        For the nonlinear MASs in equation (\ref{equ:mas}) under the event-triggered controllers in equation (\ref{fixed0})(\ref{rela0}), if the initial condition of the MASs satisfies $V(0)\leq \Omega$, the error signals $z_{i,k}$, $\tilde{c}_{i,k}$,  ${{\tilde{W}}_{i,k}}$, ${{\tilde{\Theta }}_{i}}$, ${{e}_{i,k}}$, and ${{\psi}_{i,k}}$ are all uniformly bounded. Furthermore, the errors of consensus tracking between the followers' outputs and the leader's trajectory signal can be minimized to a specified range.
    \end{theorem}

	\begin{proof}
	For the follower agents $i=1,...,N$, define the overall Lyapunov function of the MASs (\ref{equ:mas}) as 
    \vspace{-5pt}
	\begin{align}
	V = \sum\limits_{i = 1}^N {\sum\limits_{k = 1}^n {{V_{i,k}}} }  + \sum\limits_{i = 1}^N \sum\limits_{k = 1}^{n - 1} {\frac{{e_{i,k + 1}^2}}{2}}
	\end{align}

	Define the parameters $r_{i,1}$, $r_{i,2}$, $r_{i,k}$ and $r_{i,n}$ as $	{{r}_{i,1}}=-\left( {{d}_{i}}+{{b}_{i}} \right)+r_{i,1}^{*}$, $	{{r}_{i,2}}=-\frac{{{d}_{i}}+{{b}_{i}}}{2}-1+r_{i,2}^{*}$, ${{r}_{i,k}}=-\frac{3}{2}+r_{i,k}^{*}$ and ${{r}_{i,n}}=-\frac{1}{2}+r_{i,n}^{*}$ with $k=3,..., n-1$. Based on the Young's inequality, while $\left\| {{{\dot{\varpi}}}_{i,k}} \right\|\le \kappa _{i,k}^{*}\left\| {{{\tilde{\tau}}}_{i,k}} \right\|$, one has

    \vspace{-5pt}
	\begin{align}
		\dot{V}&\le\sum\limits_{i=1}^{N}\Bigg\{{\sum\limits_{k=1}^{n}{r_{i,k}^{*}}}z_{i,k}^{2}+{\sum\limits_{k=2}^{n-1}{\frac{e_{i,k+1}^{2}}{2}}}+\frac{{{d}_{i}}+{{b}_{i}}}{2}e_{i,2}^{2}\nonumber\\&+{\sum\limits_{k=1}^{n-1}{{{e}_{i,k+1}}}}{{{\dot{e}}}_{i,k+1}}+2{\sum\limits_{k=1}^{n}{\kappa_{i,k}^{*}}}\tilde{\tau}_{i,k}^{2}+\frac{{{\lambda}_{i}}}{{2{o}_{i}}}({ \Theta _{i}^{*}}^{2}-\tilde{\Theta }_{i}^{2})\nonumber\\&+{\sum\limits_{k=1}^{n}{\frac{{{h}_{i,k}}}{{2{\eta}_{i,k}}}(W_{i,k}^{\text{*}T}W_{i,k}^{\text{*}}}}-\tilde{W}_{i,k}^{T}{{{\tilde{W}}}_{i,k}})+{\sum\limits_{k=1}^{n}{{{\iota }_{i,k}}}}\nonumber\\&-\psi_{i}^{T}(({{H}_{i}}-(1+\sum\limits_{k=1}^{n}{( \kappa _{i,k}^{2}+\kappa_{i,k}^{4})}){{I}_{n}})\otimes {{I}_{m}}){{\psi}_{i}}\Bigg\}
	\end{align}
	where $\kappa_{i,k}^{*}= (3-2{{\kappa}_{i,k}})/2 $, $r_{i,k}^{*}$ and $\kappa_{i,k}^{*}$ are unknown parameters for stability analysis, with $k=1,..., n$. And $\hbar( \bullet  )$ is the eigenvalue of the given matrix. Select the matrix ${{H}_{i}}$ such that ${{\hbar}_{\min }}[ {{H}_{i}}-( 1+\sum\limits_{k=1}^{n}{(\kappa _{i,k}^{2}+\kappa_{i,k}^{4})}){{I}_{n}}]={{\wp}_{i}}/2{{\hbar }_{\max }}( {{F}_{i}} )$, where ${{\wp}_{i}}$ is a positive parameter. 
    Differentiate ${{e}_{i,2}}$ and ${{e}_{i,k+1}}$ with respect to time, one has ${{\dot{e}}_{i,2}}=-\frac{{{e}_{i,2}}}{{{m}_{i,2}}}+{\Gamma _{i,2}}$, ${{\dot{e}}_{i,k+1}}=-\frac{{{e}_{i,k+1}}}{{{m }_{i,k+1}}}+{\Gamma _{i,k+1}}$, where ${{\Gamma}_{i,2}}=-\frac{{{r}_{i,1}}{{{\dot{z}}}_{i,1}}}{{{d}_{i}}+{{b}_{i}}}+\frac{{{{\hat{\Theta }}}_{i}}}{2c_{i,1}^{2}\left( {{d}_{i}}+{{b}_{i}} \right)}{{\dot{z}}_{i,1}}E_{i,1}^{T }\left( {{T}_{i,1}} \right){{E}_{i,1}}\left( {{T}_{i,1}} \right)+\frac{{{{\dot{z}}}_{i,1}}}{2\left( {{d}_{i}}+{{b}_{i}} \right)}+\frac{{{{\hat{\Theta }}}_{i}}}{c_{i,1}^{2}\left( {{d}_{i}}+{{b}_{i}} \right)}{{z}_{i,1}}E_{i,1}^{T }\left( {{T}_{i,1}} \right){{\dot{E}}_{i,1}}\left( {{T}_{i,1}} \right)$ and ${{\Gamma }_{i,k+1}}=\frac{{{\dot{e}}_{i,k}}}{{{m }_{i,k}}}+{{q}_{i,k}}{{\dot{\psi}}_{i,1}}+\frac{{{{\hat{\Theta }}}_{i}}}{2c_{i,k}^{2}}{{\dot{z}}_{i,k}}E_{i,k}^{T }\left( {{T}_{i,k}} \right){{E}_{i,k}}\left( {{T}_{i,k}}  \right)-{{r}_{i,k}}{{\dot{z}}_{i,k}}+\frac{{{{\dot{z}}}_{i,k}}}{2}+\frac{{{{\hat{\Theta }}}_{i}}}{c_{i,k}^{2}}{{z}_{i,k}}E_{i,k}^{T }\left( {{T}_{i,k}} \right){{\dot{E}}_{i,k}}\left({{T}_{i,k}} \right)$. Inspired by~\cite{mu}, we have
    
    \vspace{-5pt}
	\begin{align}
		{{\mathbb{M}}_{i,k}}&=\Bigg\{ \sum\limits_{i=1}^{N}{\sum\limits_{k=1}^{n}{(z_{i,k}^{2}+\tilde{\tau}_{i,k}^{2}+\frac{1}{{{\eta }_{i,k}}}\tilde{W}_{i,k}^{T}{{{\tilde{W}}}_{i,k}})}}+\sum\limits_{i=1}^{N}{\frac{1}{{{o}_{i}}}\tilde{\Theta }_{i}^{2}}\nonumber\\&+\sum\limits_{i=1}^{N}{\psi _{i}^{T}({{F}_{i}}\otimes {{I}_{m}}){{\psi}_{i}} }+\sum\limits_{i=1}^{N}{\sum\limits_{k=1}^{n-1}{e_{i,k+1}^{2}}}\le 2\Omega \Bigg\}
	\end{align}
	where ${{\mathbb{M}}_{i,k}}$ is compact in ${{R}^{\dim\left( {{\mathbb{C}}_{i,k}} \right)}}$, there exists an inequality $\left\| {{\Gamma }_{i,k+1}} \right\|\le {{L}_{i,k+1}}$, where $L_{i,k+1}$ denotes a positive parameter. Applying the Young's inequality, one can get ${{\Gamma }_{i,k+1}}{{e}_{i,k+1}}\le\frac{\Xi }{2} +\frac{L_{i,k+1}^{2}e_{i,k+1}^{2}}{2\Xi }$, where $\Xi$ is a positive parameter, with $k=1,..., n-1$. Select the parameters as $\frac{1}{{{m }_{i,2}}}=\frac{{{d}_{i}}+{{b}_{i}}}{2}+\frac{L_{i,2}^{2}}{2\Xi }+m_{i,2}^{*}$ and $\frac{1}{{{m }_{i,k+1}}}=\frac{1}{2}+\frac{L_{i,k+1}^{2}}{2\Xi }+m _{i,k+1}^{*}$, where $m_{i,k+1}^{*}$ denotes an unknown positive parameter, with $k=2,...,n-1$. With $l=2,...,n$ and $k=1,...,n$, define
	\begin{align}
		\beta =&\min \left\{ -2r_{i,k}^{*},\text{ -4}\kappa _{i,k}^{*},\text{ }{{h}_{i,k}},\text{ }{{\wp}_{i}},\text{ }{{\lambda}_{i}},\text{ 2}{{m}_{i,l}^{*}} \right\}
	\end{align}
    \vspace{-10pt}
	\begin{align}
		\gamma=&\sum\limits_{i=1}^{N}{\sum\limits_{k=1}^{n}{({{\iota }_{i,k}+{\frac{{{h}_{i,k}}}{2{{\eta }_{i,k}}}W_{i,k}^{\text{*}T}W_{i,k}^{\text{*}}})}}}+ \nonumber \\
        &\sum\limits_{i=1}^{N}({\sum\limits_{k=1}^{n-1}{\frac{\Xi }{2}+{\frac{{{\lambda}_{i}}}{2{{o}_{i}}}{{\left( \Theta _{i}^{*} \right)}^{2}}}}})
	\end{align}
	
    To ensure the stability of the entire MASs, it is crucial to appropriately select parameter values while adhering to the following conditions: $-r_{i,k}^{*}>0,\text{ }-\kappa _{i,k}^{*}>0,\text{ }{{h}_{i,k}}>0,\text{ }{{\wp}_{i}}>0,\text{ }{{\lambda}_{i}}>0$ and ${{m}_{i,l}^{*}}>0$. 
    Then, we have
    \vspace{-5pt}
	\begin{align}
		\dot{V}\le -\beta V+\gamma
	\end{align}
	
	When set $\beta >\gamma/\Delta $, we get $\dot{V}<0$ on $V=\Delta $. Further, if at time $t=0$ the condition $V\le \Delta $ holds, it follows that $V\le \Delta $ for entire $t>0$. This demonstrates that the error signals ${{z}_{i,k}}$, ${{e}_{i,k}}$, ${{\tilde{\tau}}_{i,k}}$, ${{\tilde{W}}_{i,k}}$, ${{\tilde{\Theta }}_{i}}$ and ${{\psi}_{i,k}}$ are uniformly bounded. It is straightforward to derive the following:
    \vspace{-5pt}
	\begin{align}
		\frac{1}{2}{{\left\| {{\Upsilon}_{1}} \right\|}^{2}}\le V\left( t \right)\le {{e}^{-\beta t}}V\left( 0 \right)+\frac{\gamma}{\beta}\left( 1-{{e}^{-\beta t}} \right)
	\end{align}
	where ${{\Upsilon}}={{\left[ \Upsilon_1^{T },\Upsilon_2^{T },\ldots ,\Upsilon_M^{T } \right]}^{T }}$. Then, we have ${{\left\| {{\Upsilon}} \right\|}^{2}}\le 2{{e}^{-\beta t}}V\left( 0 \right)+\frac{2\gamma}{\beta}\left( 1-{{e}^{-\beta t}} \right)$.
	
	Consequently, as time progresses, all consensus tracking errors will converge to a compact set defined as  $\Im=\{{{\Upsilon}_{1}}|\left\| {{\Upsilon}_{1}} \right\|\le \sqrt{2\gamma/\beta }\}$. This means that the tracking errors can be modified and reduced to an arbitrarily small range by increasing the parameter $\beta $. From (\ref{fixed2}), one has
	\begin{align}
		{{\dot{\alpha }}_{i}}\left( t \right)={{\dot{w}}_{i,n+1}}-\frac{\bar{\pi}_{i}{{{\dot{z}}}_{i,n}}}{{{\cosh }^{2}}\left( \frac{{{z}_{i,n}}{{{\bar{\pi}}}_{i}}}{{{\mu}_{i}}} \right)}
	\end{align}
	
	According to \cite{23,jia2,jia1}, one gets
	\begin{align}
		\frac{d}{dt}\left| {{\vartheta}_{i}}\left( t \right) \right|=sign\left( {{\vartheta }_{i}}\left( t \right) \right){{\dot{\vartheta }}_{i}}\left( t \right)\le \left| {{{\dot{w }}}_{i}}\left( t \right) \right|
	\end{align}
	
	Based on the stability analysis, it is imperative that there exists a positive parameter $\varPi$ such that $| {{{\dot{w }}}_{i}}( t ) |\le \varPi$. Based on (\ref{fixed1})(\ref{fixed2}), we have ${{\vartheta }_{i}}( {{t}_{s}})=0$ and $\underset{{}}{\mathop{{{\lim }_{t\to {{t}_{s+1}}}}}}\,{{\vartheta }_{i}}(t)={{\pi}_{i}}$. Additionally, for the time interval $t\in [{{t}_{s}},{{t}_{s+1}})$, the lower bound for the inter-execution time is given by ${{t}^{*}}\ge {{\pi}_{i}}/{\varPi}$. Following the same analysis in the proof of fixed-threshold strategy, according to (\ref{rela1})(\ref{rela2}), the relative-threshold and switch-threshold strategy satisfy ${{t}^{*}}\ge ({\Delta_i|u_i(t)|+{\pi}_{i}^*})/{\varPi}$ and ${{t}^{*}}\ge {max\{{\pi}_{i},{\pi}_{i}^*\}}/{\varPi}$, respectively. Consequently, the Zeno behavior is proficiently avoided. 
	\end{proof}

	\section{Illustrative Example}\label{sec:example}
    
	This section presents an illustrative example to validate the soundness and performance of our theoretical approach. The MASs under investigation comprise one leader and four followers, where the dynamics of each agent is described by: 
	\begin{align}
		& {{{\dot{x}}}_{i,1}}={{x}_{i,2}}+{{f}_{i,1}}\left( {{{\bar{x}}}_{i,1}} \right)+{{\xi }_{i,1}} \nonumber\\
		& {{{\dot{x}}}_{i,2}}={{u}_{i}}+{{f}_{i,2}}\left( {{{\bar{x}}}_{i,2}} \right)+{{\xi }_{i,2}} \nonumber\\
		& {{y}_{i}}(t)={{x}_{i,1}}
	\end{align}
	where ${{f}_{i,1}}\left( {{{\bar{x}}}_{i,1}} \right)=0.8{{x}_{i,1}}{{e}^{-1.4x_{i,2}^{2}}}$, ${{f}_{i,2}}\left( {{{\bar{x}}}_{i,2}}\right)=-0.5x_{i,1}^{2}\cos \left( {{x}_{i,2}} \right)$, ${{\xi}_{i,1}}=0.8{{x}_{i,1}}\sin \left( {{x}_{i,2}} \right)\cos^2 \left( t \right)$ and ${{\xi }_{i,2}}=0.2{{x}_{i,2}}\cos \left( {{x}_{i,1}} \right)\cos^2 \left( t \right)$, $i=1,...,4$. The leader's trajectory signal is defined as ${{y}_{r}}=-0.5\sin(4t)\cos(2t)$.
	
	The communication topology is shown in Fig. 1. The connection matrix linking the leader to the followers is represented as $\mathcal{B}=diag(0,1,0,0)$. And the adjacency matrix $\mathcal{A}$ and the Laplacian matrix $\mathcal{L}$ are given as follows:
	\vspace{-5pt}
	$$\mathcal{A}=\left[
	\begin{array}{cccc}
		0 & 1 & 0 & 0\\
		0 & 0 & 0 & 0\\
		0 & 1 & 0 & 0\\
		1 & 0 & 0 & 0\\
	\end{array}
	\right],
	~\mathcal{L}=\left[
	\begin{array}{cccc}
		1 & -1 & 0 & 0\\
		0 & 0 & 0 & 0\\
		0 & -1 & 1 & 0\\
		-1 & 0 & 0 & 1\\
	\end{array}
	\right]$$
	
	The initial conditions of both the four followers along with their state observers are selected as follows: 
	\begin{center}
		\begin{tabular}{|c|c|}
			\hline
			${{{\bar{x}}}_{1,2}}(0)={{[0.2,\text{ }0]}^{T}}$ &  ${{{\hat{\bar{x}}}}_{1,2}}(0)={{[0.3,\text{ 1.7}]}^{T}}$  \\
			\hline 
			${{{\bar{x}}}_{2,2}}(0)={{[-0.2,\text{ 0}]}^{T}}$ & ${{{\hat{\bar{x}}}}_{2,2}}(0)={{[-0.5,\text{ 1.7}]}^{T}}$  \\
			\hline  
			${{{\bar{x}}}_{3,2}}(0)={{[0.1,\text{ 0}]}^{T}}$ & ${{{\hat{\bar{x}}}}_{3,2}}(0)={{[0,\text{ -4}]}^{T}}$ \\
			\hline
			${{{\bar{x}}}_{4,2}}(0)={{[-0.3,\text{ 0}]}^{T}}$ &  ${{{\hat{\bar{x}}}}_{4,2}}(0)={{[0,\text{ -4}]}^{T}}$ \\
			\hline 
		\end{tabular}
	\end{center}

    Following the parameter selection guidelines outlined in the stability analysis, the values of the design parameters are set as:
	for fix-threshold strategy, ${{\pi}_{i}}=2.5$, ${{\bar{\pi}}_{i}}=4$, ${{\mu}_{i}}=5.4$, for relative-threshold strategy, ${{\pi}_{i}^*}=2$, ${{\bar{\pi}}_{i}^*}=4$, $\Delta_i=0.245$, for switched-threshold strategy $G=6$, for the first order low pass filter, $m_{i} =0.005$, for other parameters, ${{h}_{i,1}}={{h}_{i,2}}=50$, ${{r}_{i,1}}={{r}_{i,2}}=-100$, ${{c}_{i,1}}={{c}_{i,2}}=100$, ${{\eta }_{i,1}}={{\eta }_{i,2}}=0.01$, ${{q}_{i,1}}=350$, ${{q}_{i,2}}=0.5$, ${{\lambda}_{i}}=120$ and ${{o}_{i}}=25$.

    \begin{figure}[ht!]
		\centering
        \includegraphics[height=6cm,width=8.5cm]{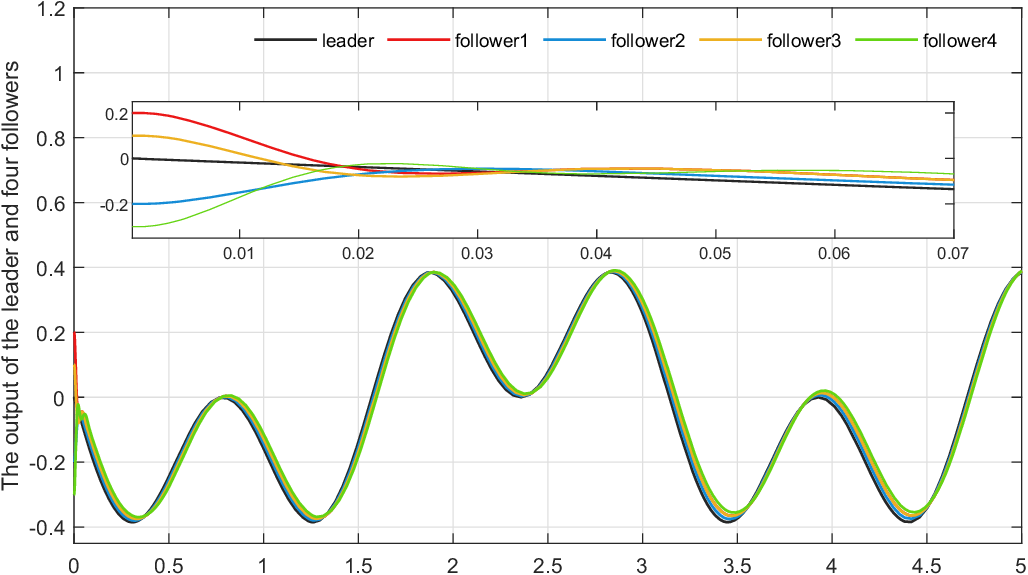}
		\caption{Consensus tracking performance.}
	\end{figure}
    
	\begin{figure}[ht!]
		\centering
		\includegraphics[height=6cm,width=8.5cm]{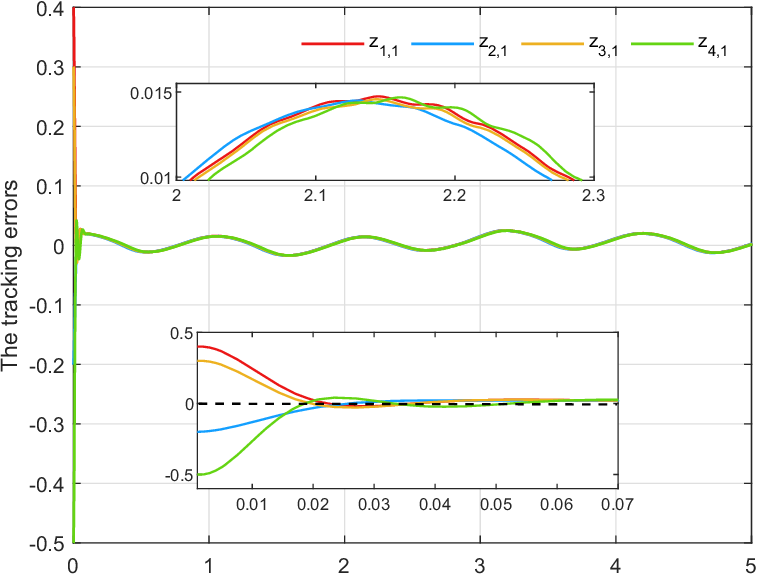}
		\caption{Trajectories of tracking errors $z_{i,1}$($i=1,2,3,4.$)}
	\end{figure}

	In Fig. 2, shortly after the process begins, all followers within the system consistently follow the leader, successfully achieving adaptive consensus tracking. Fig. 3 illustrates that the tracking error has converged to zero. The total triggering number for the sample-data strategy is 5,000 times, and the triggering numbers for the three threshold strategies are detailed in Table I. Figs. 4 shows the interval of events respectively triggered by the three strategies. The statistics demonstrates that the highest number of triggers arises from the fixed-threshold strategy. Under the relative-threshold strategy, the number of triggers is the lowest, while under the switched-threshold strategy, the number of triggers falls and both types play an indispensable role in this strategy.

    \begin{table}[ht!]
		\centering
		\caption{\small{Comparison of three event-triggered strategies.}}
		\label{tab:performance}
		\begin{tabular}{cccc}  
			\toprule   
			$\text{\text{ }}$   & $\text{Fixed-threshold}$ & $\text{Switch-threshold}$ & $\text{Relative-threshold}$\\
			\midrule
			$\text{Follower 1}$  & $364$          & $344(255+89)$      & $310$        \\
			$\text{Follower 2}$ & $296$           & $281(220+61)$       & $277$      \\
			$\text{Follower 3}$ & $358$          & $342(255+87)$      & $308$   \\
			$\text{Follower 4}$ & $453$          & $420(306+114)$      & $380$   \\
			\bottomrule
		\end{tabular}
	\end{table}

	\begin{figure*}[h]	
		\begin{minipage}{0.33\linewidth}
			\vspace{3pt}
			\centerline{\includegraphics[height=6.5cm,width=6.5cm]{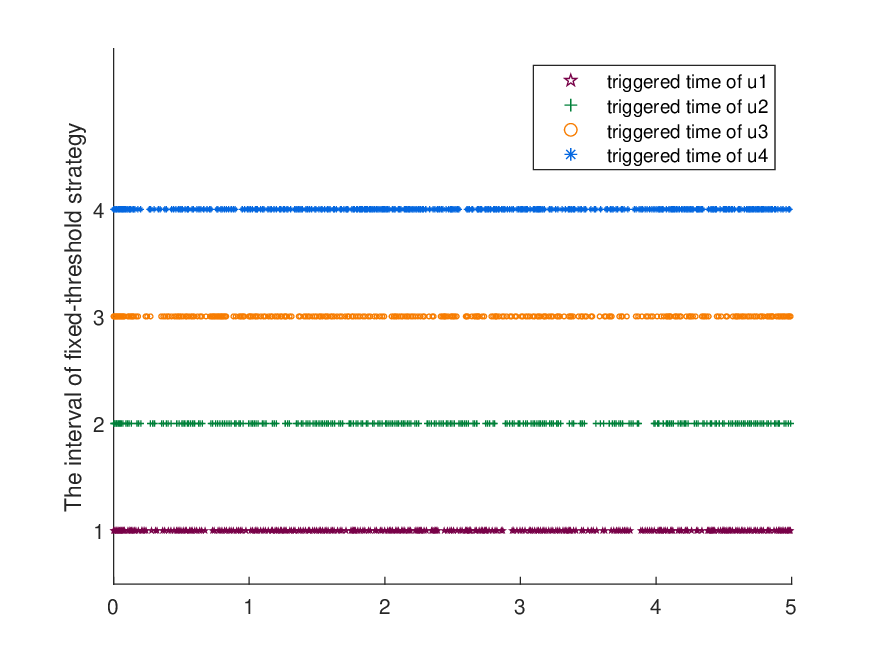}}
            \vspace{-10pt}
			\centerline{\small Fixed-threshold strategy}
		\end{minipage}
		\begin{minipage}{0.33\linewidth}
			\vspace{3pt}
			\centerline{\includegraphics[height=6.5cm,width=6.5cm]{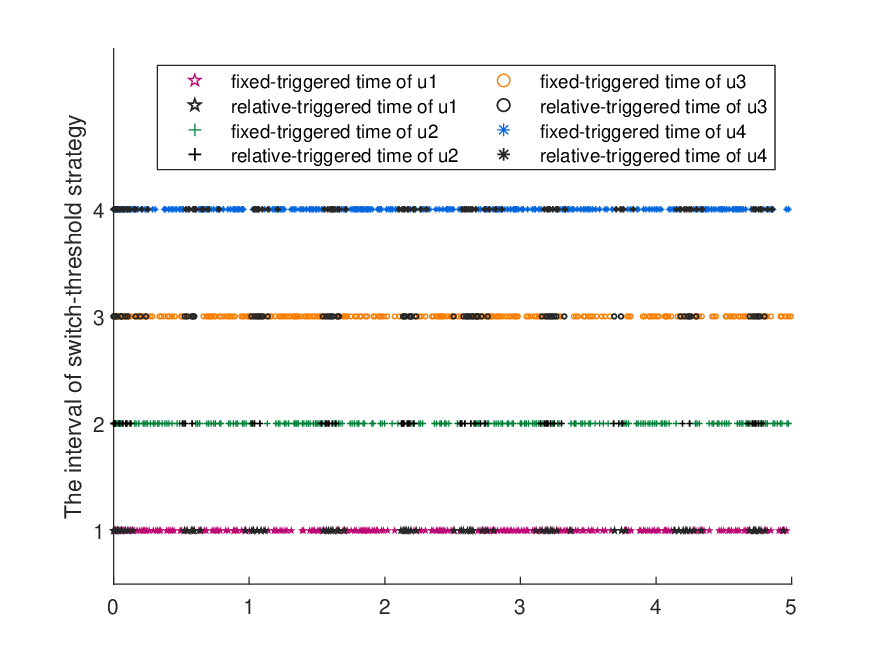}}
			 \vspace{-10pt}
			\centerline{\small Switched-threshold strategy}
		\end{minipage}
		\begin{minipage}{0.33\linewidth}
			\vspace{3pt}
			\centerline{\includegraphics[height=6.5cm,width=6.5cm]{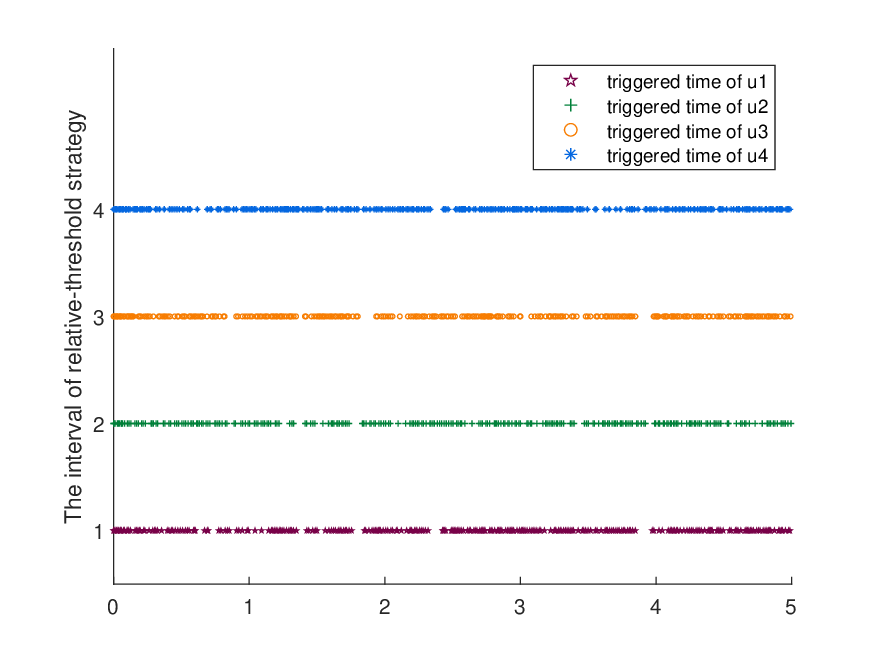}}
			 \vspace{-10pt}
			\centerline{\small Relative-threshold strategy}
		\end{minipage}
		\caption{Release instants and interval and three event-triggered control strategies.}
	\end{figure*}

	\section{Conclusion}\label{sec:conclusion}

	In this paper, we propose an observer-based control framework that addresses the consensus tracking problem for nonlinear multi-agent systems (MASs). We utilize Radial Basis Function (RBF) neural networks to approximate the unknown dynamics, which is then leveraged to build observers to measure unidentified system states and external perturbations. Then we present three event-triggered control strategies and analyze their stability using Lyapunov methods. Our approach successfully mitigates complexity issues through the integration of a filtering mechanism at each design stage. Simulation results demonstrate that the fixed-threshold strategy enhances system performance and minimizes the number of triggering instances. Meanwhile, the relative-threshold strategy further optimizes resource conservation. The switched-threshold strategy achieves a commendable balance between performance improvement and resource utilization. In our future research we plan to extend the proposed framework to address optimal event-triggered control problems for stochastic nonlinear MASs.


	\newpage

	\vfill
	

\begin{thebibliography}{1}
		\bibliographystyle{IEEEtran}
		\bibitem{intro1}
		B. V. Solanki, A. Raghurajan, K. Bhattacharya and C. A. Cañizares, "Including Smart Loads for Optimal Demand Response in Integrated Energy Management Systems for Isolated Microgrids," \textit{IEEE Transactions on Smart Grid}, vol. 8, no. 4, pp. 1739-1748, 2017.
		\bibitem{intro2}
		F. Perez, G. Damm, C. M. Verrelli and P. F. Ribeiro, "Adaptive Virtual Inertia Control for Stable Microgrid Operation Including Ancillary Services Support," \textit{IEEE Transactions Control Systems Technology}, vol. 31, no. 4, pp. 1552-1564, 2023.
		\bibitem{intro3}
        Y. Zhang, R. Zhong, and H. Yu, "Mitigating stop-and-go traffic congestion with operator learning," \textit{Transportation Research Part C Emerging Technology}, vol. 170, no. 104928, 2025.
		\bibitem{intro3.5}
        Y. Zhang, H. Yu, J. Auriol, and M. Pereira, "Mean-square exponential stabilization of mixed-autonomy traffic PDE system," \textit{Automatica}, vol. 170, no. 111859, 2024.
		\bibitem{intro4}
		X. Ge, Q. -L. Han, J. Wang and X. -M. Zhang, "Scalable and Resilient Platooning Control of Cooperative Automated Vehicles," \textit{IEEE Trans. on Vehicular Technology}, vol. 71, no. 4, pp. 3595-3608, 2022.
		\bibitem{intro5}
		S. Zhang, Z. W. Liu, Y. Zhai, Y. Zhao, and G. Wen, "Constant-stepsize distributed optimization algorithm with malicious nodes," \textit{2021 International Conf. on Neuromorphic Computing}, pp. 177-182, 2021.
		\bibitem{intro6}
		L. Wang, X. Wang, and Z. Wang, "Event-triggered optimal tracking control for strict-feedback nonlinear systems with non-affine nonlinear faults," \textit{Nonlinear Dynamics}, vol. 112, no. 17, pp. 15413-15426, 2024.
        \bibitem{intro7}
        S. Zhang, Z. Liu, G. Wen, and Y. Wang, "Accelerated distributed optimization algorithm with malicious nodes," \textit{IEEE Transactions on Network Science and Engineering}, vol. 11, no. 2, pp. 2238--2248, 2023.

		\bibitem{eve4}
		P. Tabuada, "Event-Triggered Real-Time Scheduling of Stabilizing Control Tasks," \textit{IEEE Transactions on Automatic Control}, vol. 52, no. 9, pp. 1680-1685, 2007, 
		\bibitem{eve5}
		A. Anta and P. Tabuada, "To Sample or not to Sample: Self-Triggered Control for Nonlinear Systems," \textit{IEEE Transactions on Automatic Control}, vol. 55, no. 9, pp. 2030-2042, 2010.
		\bibitem{eve1}
		K. J. Åström and B. Bernhardsson, "Comparison of periodic and event based sampling for first-order stochastic systems," \textit{IFAC Proceedings Volumes}, vol. 32, no. 2, pp. 5006-5011, Jul. 1999.
		\bibitem{eve2}
		D. Theodosis and D. V. Dimarogonas, "Event-Triggered Control of Nonlinear Systems With Updating Threshold," \textit{IEEE Control Systems Letters}, vol. 3, no. 3, pp. 655-660, 2019.
		\bibitem{eve3}
		X. Wang, Y. Zhou, T. Huang and P. Chakrabarti, "Event-Triggered Adaptive Fault-Tolerant Control for a Class of Nonlinear Multiagent Systems With Sensor and Actuator Faults," \textit{IEEE Trans Circuits Systems I Regular Papart}, vol. 69, no. 10, pp. 4203-4214, 2022.
		\bibitem{eve3.1}
		P. Elena, P. Romain, A. Daniele, N. Dragan and H. W. Maurice, "Decentralized event-triggered estimation of nonlinear systems,"\emph{Automatica}, vol. 160, pp. 111414, 2024.
		\bibitem{eve3.2}
		Z. M. Wang, "Hybrid Event-triggered Control of Nonlinear System with Full State Constraints and Disturbance," \textit{2024 36th Chinese Control and Decision Conference}, pp. 2122-2127, 2024.
        \bibitem{eve3.3}
		R. Postoyan, A. Anta, W. P. M. H. Heemels, P. Tabuada and D. Nešić, "Periodic event-triggered control for nonlinear systems," \textit{52nd IEEE Conference on Decision and Control}, pp. 7397-7402, 2013.
		\bibitem{pang}
        N. Pang, L. Y. Huang, B. T. Dong, H. T. Chen, Z. H. Jia and W. D. Zhang, "Safe Policy Optimization With Stretchable Penalties," \textit{Authorea Preprints}, DOI:10.36227/techrxiv.173092241.15952706/v1 2024. 
		\bibitem{chufa}
		L. T. Xing, C. Y. Wen, Z. T. Liu, H. Y. Su, and J. P. Cai, "Event-Triggered Adaptive Control for a Class of Uncertain Nonlinear Systems," \textit{IEEE Trans. Autom. Contrd}, vol. 62, no. 4, pp. 2071-2076, 2017.
        \bibitem{jiang2024}
        G. Jiang, Y. Wang, Y. Li, N. S. Moosavi, and P. Hui, "Blending social interaction realms: Harmonizing online and offline interactions through augmented reality," in \textit{17th International Symposium on Visual Information Communication and Interaction}, pp. 1–8, 2024.
		\bibitem{lemma1}
		C. L. P. Chen, G. X. Wen, Y. J. Liu, and Z. Liu, "Observer-based adaptive backstepping consensus tracking control for high-Order nonlinear semi-strict-feedback multiagent systems,"\emph{IEEE Transactions on Cybernetics}, vol. 46, no. 7, pp. 1591--1601, 2016.
		\bibitem{lemma2}
		Z. M. Wang, X. Wang and N. Pang, "Adaptive Fixed-Time Control for Full State-Constrained Nonlinear Systems: Switched-Self-Triggered Case" \textit{IEEE Trans. Circuits Systems II Express Briefs}, vol. 71, no. 2, pp. 752--756, 2024.
		\bibitem{tnnls}
		N. Pang, X. Wang and Z. M. Wang, "Observer-Based Event-Triggered Adaptive Control for Nonlinear Multiagent Systems With Unknown States and Disturbances," \textit{IEEE Transactions on Neural Networks and Learning Systems}, vol. 34, no. 9, pp. 6663-6669, Sept. 2023.
		\bibitem{mu}
		S. J. Yoo, "Distributed consensus tracking for multiple uncertain nonlinear strict-feedback systems under a directed graph," \textit{IEEE Trans. on Neural Netw. Learn. Syst}, vol. 24, no. 4, pp. 666–672,  2013.

		\bibitem{rela1}
		L. Liu and X. Li, "Event-Triggered Tracking Control for Active Seat Suspension Systems With Time-Varying Full-State Constraints," \textit{IEEE Trans. on Syst. Man Cybern. Syst.}, vol. 52, no. 1, pp. 582-590, 2022.
		\bibitem{rela2}
		Z. M. Wang, H. Wang, X. Wang, N. Pang and Q. Shi, "Event-Triggered Adaptive Neural Control for Full State-Constrained Nonlinear Systems with Unknown Disturbances" \textit{Cognitive Computation}, vol. 16, no. 2, pp. 717--726, 2023. 
		\bibitem{23} 
		T. Henningsson, E. Johannesson, and A. Cervin, "Sporadic event-based control of first-order linear stochastic systems,"\emph{Automatica}, vol. 44, no. 11, pp. 2890--2895, 2008.
        \bibitem{jia2}
        N. Pang, X. Wang, and Z. Wang, "Event-triggered adaptive control of nonlinear systems with dynamic uncertainties: The switching threshold case," \textit{ IEEE Transactions on Circuits and Systems II: Express Briefs}, vol. 69, no. 8, pp. 3540-3544, 2022.
        \bibitem{jia1}
            G. Zhang, X. Wang, Z. Wang, and N. Pang, "Optimized backstepping tracking control using reinforcement learning for strict-feedback nonlinear systems with monotone tube performance boundaries," \textit{International Journal of Control}, pp. 1–13, 2025.







        
	\end{thebibliography}
\end{document}